\documentclass[12pt,preprint]{aastex}

\usepackage{epsfig}
\usepackage{graphics}
\usepackage{graphicx}
\usepackage{amsmath}

\slugcomment{KSUPT -- 08/6  \quad  October 2008}

\begin{document}

\title{Constraints on dark energy models from radial baryon acoustic scale measurements}

\author{Lado Samushia\altaffilmark{1,2} and Bharat Ratra\altaffilmark{1}}

\altaffiltext{1}{Department of Physics, Kansas State University, 116 Cardwell Hall, Manhattan, KS 66506, \\
lado@phys.ksu.edu, ratra@phys.ksu.edu.}
\altaffiltext{2}{National Abastumani Astrophysical Observatory, 2A Kazbegi Ave, GE-0160 Tbilisi, Georgia.}

\begin{abstract}
We use the radial baryon acoustic oscillation (BAO) measurements of \citet{gaztanaga08a} to constrain parameters of dark energy models. These constraints are comparable with constraints from other ``non-radial" BAO data. The radial BAO data are consistent with the time-independent cosmological constant model but do not rule out time-varying dark energy. When we combine radial BAO and the \citet{kowalski08} Union type Ia supernova data we get very tight constraints on dark energy.
\end{abstract}


\section{Introduction}

Recent type Ia supernova data \citep{kowalski08,rubin08,sahni08} confirm, at high significance, that the cosmological expansion is currently accelerating. If we assume that general relativity is valid on cosmological length scales, this cosmological acceleration requires that the Universe's current energy density budget is dominated by far by an approximately spatially uniform component -- dark energy -- with negative pressure $p<-\rho/3$, where $\rho$ is the dark energy density. 

The most economic and the oldest form of dark energy is Einstein's cosmological constant $\Lambda$ \citep{pee84} which is time independent and has an equation of state $p_{\Lambda}=-\rho_{\Lambda}$. The time-independent cosmological constant model ($\Lambda$CDM) provides a reasonably good fit to most current cosmological data \citep[see, e.g.,][]{ratra08,frieman08}, but despite this success a lot of alternative models of dark energy have been proposed over the years. One reason for this is that cosmological data can not yet tightly constrain the various options currently under debate, although it is thought that in next decade a large amount of more precise new data will tightly constrain or measure departures from the now standard $\Lambda$CDM model \citep[see, e.g.,][]{pod01a,wang07b,barnard08,tang08}.\footnote{The $\Lambda$CDM model is based on the cold dark matter (CDM) structure formation model which might have trouble accounting for some observations (for a summary of some of the problems see \citealt{pee03}).} The second reason is that a cosmological constant is not straightforward to understand on a more fundamental level; in particular, it is  difficult to accept as fundamental the needed new energy scale of a few ${\rm meV}$.

In some scalar field dark energy models a nonlinear attractor solution ensures that this scale of a few ${\rm meV}$ is not fundamental, rather it follows from a much higher energy scale as the scalar field energy density decreases during the cosmological expansion (\citealt{pee88,rat88}). In this paper we also consider this $\phi$CDM model in which the dark energy is modeled as a slowly rolling scalar field $\phi$ with self-interaction potential energy density $V(\phi)\propto \phi^{-\alpha}$, where $\alpha>0$. A number of other models with time-varying dark energy have been proposed.\footnote{Modifications of general relativity on cosmological length scales that might do away with dark energy have also been discussed \citep[see, e.g.,][]{wei08,tsujikawa08,capozziello08,sotiriou08,bamba08}. In the models we consider in this paper, dark energy affects other fields only through gravity. For dark energy models with other couplings see \citet{virgani08}, \citet{wuzhangsp08}, \citet{jesus08}, \citet{antusch08}, \citet{wuzhang08}, and references therein. For other dark energy models see \citet{zhang08}, \citet{alcaniz08}, \citet{setare08}, \citet{basilakos08}, \citet{shapiro08}, and references therein.} We also consider the XCDM parametrization which is often used to describe time-varying dark energy models.\footnote{The XCDM parametrization cannot describe time-varying $\phi$CDM dark energy at late times (see, e.g., \citealt{rat91}).} In XCDM dark energy is assumed to be a perfect fluid with effective equation of state $p_{\rm x}=\omega_{\rm x}\rho_{\rm x}$, where $\omega_{\rm x}$ is a number less than $-1/3$.

Given a model it is possible to compute quantities such as the Hubble parameter $H(z)$ and the angular diameter distance $d_A(z)$ as a function of redshift $z$. Since these quantities are observable (or observational results depend on them), given perfect data it is possible in principle to compare observational results with theoretical predictions and determine which dark energy model provides a better fit to the data. There are a few difficulties however. Theoretical predictions not only depend on dark energy model parameters, they also can depend on a number of other cosmological parameters, such as the energy density of non-relativistic matter or baryonic matter or radiation, the energy density perturbation spectral index, the total mass of neutrinos, the Hubble constant, etc. While some of these parameters are directly measurable, some of them have to be simultaneously determined from the data. This leads to degeneracies in cosmological parameter space.

The other issue is possible systematic errors in cosmological observations, which are very difficult to trace. It is therefore reassuring that the supernova data indication of dark energy is confirmed by cosmic microwave background (CMB) anisotropy data. More precisely, assuming the CDM model of structure formation and assuming that dark energy does not evolve in time, CMB anisotropy data are consistent with negligible space curvature \citep[see, e.g.,][]{pod01b,page03,doran07,dunkley08,komatsu08}, so in conjunction with low measured non-relativistic matter density \citep[see, e.g.,][]{che03b} CMB anisotropy data also demand dark energy. 

To convincingly remove degeneracies and understand and cancel the effect of unwanted systematic errors it's necessary to have many independent cosmological tests. Other recently discussed cosmological probes include the angular size of radio sources and quasars as a function of redshift (see, e.g., \citealt{chen03a,pod03,dal07,santos08}), strong gravitational lensing (see, e.g., \citealt{lee07,oguri08,zhang07,zhu08}), weak gravitational lensing (see, e.g., \citealt{dore07,lavacca08,hoekstra08,schmidt08}), measurements of the Hubble parameter as a function of redshift (see, e.g., \citealt{samushia06,lin08,dev08a,fernandez-martinez08}), galaxy cluster gas mass fraction versus redshift data (see, e.g., \citealt{allen04,che04,allen08,sen08}), and the growth of large-scale structure \citep[see, e.g.,][]{basilakosetal08,waizmann08,mainini08,abramo08}.

A measurement of the large-scale structure baryon acoustic oscillation (BAO) peak length scale provides another possible cosmological test. This was first measured a few years ago in the two-point correlation function of luminous red galaxies in the SDSS catalog at $z=0.35$ (\citealt{eisensteinetal}, also see \citealt{hutsi06}) and from 2dFGRS data \citep{cole05}, and then later in the joint SDSS and 2dF catalogs at $z=0.24$ and $0.43$ \citep{percivaletal07a}.\footnote{See \citet{blake07}, \citet{padmanabhan07}, and \citet{percivaletal07b} for more recent discussion of the observational situation. See \citet{blake03}, \citet{linder03}, \citet{seo03}, \citet{hu03}, and references therein, for discussions of the BAO peak length scale ``standard ruler'' cosmological test.} One attractive feature of the BAO measurements is that they do not explicitly depend on the value of Hubble constant. These BAO measurements were used in the original papers to constrain parameters of the $\Lambda$CDM and XCDM models. See \citet{ishida08}, \citet{lazkoz07a}, \citet{santosjesus08}, and \citet{samushia08b}, and references therein, for recent discussions of BAO data constraints on these and other models. 

The \citet{percivaletal07a} measurement was made at only two redshifts and by itself does not provide a robust and reliable test of dark energy models, but in combination with other data it does provide useful constraints on cosmological parameters. However, a number of surveys are planned in next few years that will measure the BAO scale accurately and at a variety of redshifts up to $z=1.2$. This upcoming BAO data, especially when combined with other data, will prove very useful in tightly constraining dark energy parameters \citep[see, e.g.,][]{wang08}.

Recently \citet{gaztanaga08a} argued that these measurements of the BAO scale were essentially measurements orthogonal to the line of sight and so statistically independent from a line of sight measurement of the BAO scale, even if the same galaxy catalog is used for both measurements. \citet{gaztanaga08a} used the SDSS data to compute the line of sight or radial BAO scale in redshift space for two ranges of redshift and \citet{gaztanaga08b}  showed the resulting constraints on the spatially-flat XCDM model. These constrains from the radial BAO scale data are quite similar to the constraints derived from earlier ``non-radial'' BAO scale measurements. In this paper we extend the analysis to the $\Lambda$CDM and $\phi$CDM models of dark energy.\footnote{\citet{gaztanaga08b} quote limits on the $\Lambda$CDM parameters but do not show constraint contours for $\Lambda$CDM.} The data are consistent with spatially-flat $\Lambda$CDM. However, these current radial BAO measurements (like current non-radial BAO measurements) can not tightly constrain time-varying dark energy, although the situation is anticipated to improve in the next few years. 

We also derive constraints on these models from a combined analysis of the radial BAO data and the \citet{kowalski08} Union type Ia supernova data. Since the radial BAO and type Ia supernova data constraints are almost orthogonal, the constraints from the combined data are significantly tighter than those from either individual data sets. The combined constraints favor a close to spatially-flat $\Lambda$CDM model (more so than the type Ia supernova data), but do not yet completely rule out time-varying dark energy.

Our paper is organized as follows. In Sec.\ 2 we describe the radial BAO measurements we use. In Sec.\ 3 we explain how we derive constraints on different dark energy models from radial BAO and type Ia supernova data. We present and discuss our results in Sec.\ 4.

\section{Radial baryon acoustic scale}

In a spherically symmetric Universe the two-point correlation function is a function of two variables, $\xi=\xi(\sigma,\pi)$, where $\sigma$ is the separation along the line of sight and $\pi$ is the separation on the sky. It can also be expressed as a function of absolute separation $r=\sqrt{\sigma^2+\pi^2}$ and the cosine of the angle between the line of sight and the direction of separation, $\mu=\pi/r$. The correlation function can then be decomposed into multipole moments

\begin{equation}
\xi_l(r)=\int_{-1}^{+1}{\xi(r,\mu)P_l(\mu)d\mu},
\end{equation}

\noindent
where $P_l$ is the $l^{\rm {th}}$ order Legendre polynomial. Multipole moments of different orders can be related to each other if one has a complete theory of linear and nonlinear evolution. Although high multipoles that describe the ``shape'' of baryon acoustic oscillation imprints on the matter distribution are very difficult to measure in practice, theoretically they are independent of the monopole and could provide additional structure formation tests. 

Initial work considered only the averaged over direction monopole part of the correlation function,

\begin{equation}
\xi_0(r)=\frac{1}{2}\int_{-1}^{+1}{\xi(r,\mu)d\mu},
\end{equation}

\noindent
and found a BAO peak signal at a comoving distance of $r\approx110 h^{-1}\rm{ Mpc}$ ($h$ is the Hubble constant $H_0$ in units of 100 $\rm {km s^{-1} Mpc^{-1}}$). This measurement was however mostly transverse to the line of sight direction $\pi$; the weight of separation along the line of sight  contributes less then $1\%$ \citep{gaztanaga08a}. Consequently, it is fair to assume that the radial baryon acoustic peak scale measurement in the line of sight direction from $\xi(\sigma)$ is statistically independent from that measured from $\xi(r)\approx\xi(\pi)$, even if the same galaxy sample is used for both measurements.

\citet{gaztanaga08a} recently used SDSS data to measure the radial baryon acoustic scale in two redshift ranges $z\sim0.15-0.30$ with radial BAO peak scale $\Delta z=0.0407 \pm 0.0014$ and $z\sim0.40-0.47$ with $\Delta z=0.0442 \pm 0.0016$ (both one standard deviation errors). Theoretically the radial BAO peak scale is given by

\begin{equation}
\Delta z=H(z)r_s(z_{\rm d})/c
\end{equation}

\noindent
where $H(z)$ is the Hubble parameter at redshift $z$, $r_s(z_{\rm d})$ is the sound horizon size at the drag redshift $z_{\rm d}$, at which baryons were released from photons, and $c$ is the speed of light. $H(z)$ can be easily computed in a given cosmological model and depends on model parameters such as the non-relativistic matter density and the time dependence of dark energy. 

To compute $r_s$ \citet{gaztanaga08b} use two different methods. One is to use the ratio $l_s$ between the distance to the last-scattering surface and $r_s$ measured by CMB anisotropy experiments and compute the sound horizon at photon decoupling from
 
 \begin{equation}
 \label{horizon1}
 r_s(z_*)=\frac{\pi(1+z_*)d_A(z_*)}{l_s}.
 \end{equation}

\noindent
Here $z_*$ is the redshift at the photon decoupling and $d_A$ is the angular diameter distance. Alternatively, one can use priors on the fractional energy density parameters of baryonic matter, $\Omega_{\rm b}$, nonrelativistic matter, $\Omega_{\rm m}$, and relativistic matter, $\Omega_{\rm r}$, from, e.g., CMB anisotropy measurements, and compute the sound horizon at the drag redshift from

\begin{equation}
\label{horizon2}
r_s(z_{\rm d})=\frac{c}{H_0\sqrt{3\Omega_{\rm m}}}\int_{0}^{a(z_{\rm d})}{\frac{da}{\sqrt{(a+1.69\Omega_{\rm r}/\Omega_{\rm m})(1+a0.75\Omega_{\rm b}/\Omega_{\rm r})}}}.
\end{equation}
 
\noindent

Both options have similar drawbacks. One has to assume priors on ``nuisance'' parameters like $l_s$ or various energy densities. CMB anisotropy measurements themselves have measurement errors that must be accounted for, otherwise the errors on the estimates of dark energy model parameters of interest will be underestimated. Also, the best fit values for nuisance parameters given by CMB anisotropy data are in general different for every cosmological model and also depend on model parameter values. To be fully consistent when using priors one would have to reanalyze CMB experiments for each cosmological model (and model parameter value) instead of using a single set of values for $l_s$, $\Omega_{\rm b}$, $\Omega_{\rm m}$, and $\Omega_{\rm r}$.  

At present, however, the BAO scale is measured only in two redshift ranges and does not provide very tight parameter constraints compared to other observational tests. Hence, as long as we are interested in preliminary constraints on dark energy from BAO scale measurements we may use the simplified approach of \citet{gaztanaga08b}, keeping in mind that when more and better quality BAO scale measurements become available a more complete, careful, and time-consuming analysis will be warranted.

\section{Computation}

We compare the two measured radial BAO peak scales \citep{gaztanaga08a} with the predictions of three dark energy models. The models we consider are standard $\Lambda$CDM, the XCDM parametrization of the dark energy's equation of state, and the $\phi$CDM model with an inverse power law potential energy density $V(\phi)\propto \phi^{-\alpha}$. In these dark energy models, at late times, we can compute the redshift-dependent Hubble parameter from

\begin{align}
&H(z)=H_0\sqrt{\Omega_{\rm m}(1+z)^3+\Omega_\Lambda+(1-\Omega_{\rm m}-\Omega_\Lambda)(1+z)^2}\qquad &(\Lambda \rm {CDM}),\\
&H(z)=H_0\sqrt{\Omega_{\rm m}(1+z)^3+(1-\Omega_{\rm m})(1+z)^{3(1+\omega_{\rm x})}}\qquad &(\rm {XCDM}),\\
&H(z)=H_0\sqrt{\Omega_{\rm m}(1+z)^3+\Omega_\phi(\alpha,z)}\qquad &(\phi \rm {CDM}).
\end{align}

\noindent
In the $\phi\rm{CDM}$ model the energy density of the scalar field has to be computed as a function of the redshift and $\alpha$ parameter by numerically solving the equations of motion.\footnote{For a discussion of the scalar field dynamics see, e.g., \citet{podariu00}.}

In all three models the background evolution is described by two parameters. One is the nonrelativistic matter fractional energy density parameter $\Omega_{\rm m}$ and the other one is a parameter $p$ that characterizes the dark energy. For the $\Lambda$CDM model $p$ is the cosmological constant fractional energy density parameter $\Omega_\Lambda$, in the XCDM parametrization it is the equation of state parameter $\omega_{\rm x}$, and in the $\phi \rm{CDM}$ model it is $\alpha$ which describes the steepness of the scalar field self-interaction potential. In this paper we consider only spatially-flat XCDM and $\phi$CDM models, while in the $\Lambda$CDM case spatial curvature is allowed to vary, with the space curvature fractional energy density parameter $\Omega_{\rm k}=1-\Omega_{\rm m}-\Omega_\Lambda$. The angular diameter distance is determined in terms of the Hubble parameter through

\begin{align}
&d_A(z)=\frac{1}{\sqrt{\Omega_{\rm k}}H_0(1+z)}\sin{\left[\sqrt{\Omega_{\rm k}}H_0\int_0^z{\frac{dz'}{H(z')}}\right]}\quad &(\Omega_{\rm k}>0),\\
&d_A(z)=\frac{1}{(1+z)}\int_0^z{\frac{dz'}{H(z')}}\quad &(\Omega_{\rm k}=0),\\
&d_A(z)=\frac{1}{\sqrt{-\Omega_{\rm k}}H_0(1+z)}\sinh{\left[\sqrt{-\Omega_{\rm k}}H_0\int_0^z{\frac{dz'}{H(z')}}\right]}\quad &(\Omega_{\rm k}<0).
\end{align}

We use both methods proposed in \citet{gaztanaga08b} to estimate the sound horizon at recombination.  We compute theoretically predicted model values of the radial baryon acoustic scale in redshift space, $\Delta z_{\rm {th}}(\Omega_{\rm m}, p, \boldsymbol\nu)$, at redshift $z_1=0.24$ and $z_2=0.43$,
where by $\boldsymbol\nu$ we denote the ``nuisance'' parameters.

We assume that the measurements at $z=0.24$ and $0.43$ are independent and that the errors are Gaussianly distributed. To constrain model parameters we compute

\begin{equation}
\label{chi}
\chi^2(\Omega_{\rm m}, p, \boldsymbol\nu)=\sum_{i=1,2}{(\Delta z_{{\rm th},i}(\Omega_{\rm m}, p, \boldsymbol\nu)-\Delta z_{{\rm obs},i})^2/\sigma_{\Delta z, i}^2},
\end{equation}
\noindent
where the two observed values, $\Delta z_{{\rm obs},i}\pm \sigma_{\Delta z,i}$, are listed above Eq.\ (3) and define a likelihood function

\begin{equation}
\label{lik}
\mathcal{L}_{\rm BAO}(\Omega_{\rm m}, p, \boldsymbol\nu)\propto\exp(-\chi(\Omega_{\rm m}, p, \boldsymbol\nu)/2).
\end{equation}

We integrate over nuisance parameters to get a two dimensional likelihood for the cosmological parameters of interest $\mathcal{L}_{\rm BAO}(\Omega_{\rm m}, p)=\int{\mathcal{L}_{\rm BAO}(\Omega_{\rm m}, p, \boldsymbol\nu)\mathcal{L}(\boldsymbol\nu)d\boldsymbol\nu}$. For the prior likelihood of nuisance parameters we use gaussian distribution functions with WMAP 5-year recommended means and variances $l_s=302\pm0.87$, $z_s=1090\pm1$ (with $40\%$ positive correlation between two measurements), $\Omega_{\rm b}h^2=0.0227\pm0.0066$, $\Omega_{\rm m}h^2=0.133\pm0.0064$, and $\Omega_{\rm r}=2.45\times10^{-5}$ \citep{dunkley08}. For each dark energy model we define the best-fit values of parameters as a pair ($\Omega_{\rm m}$, $p$) that maximizes the two-dimensional likelihood function $\mathcal{L}_{\rm BAO}(\Omega_{\rm m}, p)$. The 1, 2, and 3$\sigma$ confidence level contours are defined as the sets of parameters ($\Omega_{\rm m}$, $p$) for which $\mathcal{L}_{\rm BAO}(\Omega_{\rm m}, p)$ is less than its maximum value by multiplicative factors of $\exp(-2.30/2)$, $\exp(-6.18/2)$, and $\exp(-11.83/2)$, respectively.  In Figs.\ 1--3 we show two-dimensional constraints on cosmological parameters from radial BAO data for each of three dark energy models and two sets of priors.

Since radial BAO measurements alone, at the moment, can not tightly constrain cosmological parameters we combined them with the \citet{kowalski08} Union type Ia supernova data. In our derivation of the type Ia supernova constraints we closely follow \citet{kowalski08}. We define a likelihood function for cosmological parameters, $\mathcal{L}_{\rm SN}(\Omega_{\rm m}, p, H_0)$, for each dark energy model. We marginalize over $H_0$ with a flat noninformative prior in the range $40 < H_0/(\rm km\ s^{-1}\ Mpc^{-1}) < 100$ to get a two-dimensional likelihood $\mathcal{L}_{\rm SN}(\Omega_{\rm m}, p)$. The constraints on the $\phi$CDM model from the supernova data are shown in Fig.\ 4.\footnote{For the constraints on $\Lambda$CDM and XCDM models from the supernova ``Union'' data set see Fig.\ 12 in \citet{kowalski08}.} Since radial BAO and supernova measurements are independent, we multiply the individual likelihoods to get a joint likelihood function $\mathcal{L}(\Omega_{\rm m}, p)=\mathcal{L}_{\rm BAO}(\Omega_{\rm m}, p)\mathcal{L}_{\rm SN}(\Omega_{\rm m}, p)$. We show constraints derived from the joint likelihood function in Figs.\ 5--7.

From the joint two-dimensional likelihood function we define the marginal one-dimensional likelihood functions for parameters $\Omega_{\rm m}$ and $p$ (assuming uniform priors on both parameters in the ranges $\Omega_{\rm m}\in (0.0,1.0)$, $\Omega_\Lambda\in(0.0,1.0)$, $\omega_{\rm x}\in(-2.0,0.0)$, and $\alpha\in(0.0,5.0)$) through

\begin{align}
\mathcal{L}(\Omega_{\rm m})=\int \mathcal{L}(\Omega_{\rm m},p) dp, \\
\mathcal{L}(p)=\int \mathcal{L}(\Omega_{\rm m},p) d\Omega_{\rm m}.
\end{align}

\noindent
In Fig.\ 8 we show one-dimensional likelihood functions for $\Omega_{\rm m}$ and $p$ for the three dark energy models. From these one-dimensional likelihood functions, in all three models, we define the best-fit value of parameters as the values that maximize the likelihood, with highest posterior density 1$\sigma$ confidence level intervals as the value of parameters for which $\int_{x\in 1\sigma}dx\, \mathcal{L}(x)/\int_{\forall x}dx\, \mathcal{L}(x)=0.68$ and with $\mathcal{L}(x)$ higher everywhere inside the interval than outside. The values of one-dimensional best-fit parameters and corresponding 1$\sigma$ confidence level intervals are listed in Table\ 1.

\section{Results and discussion}

Figure\ 1 shows the \citet{gaztanaga08a} radial BAO scale constraints on the $\Lambda$CDM model. When we use the WMAP measured value of the ratio $l_s$ (thick lines) the model is constrained to be very close to the spatially-flat case. When we use the WMAP measured value of the sound horizon at recombination $r_s$ (thin lines) the constraints are weaker, the spatial curvature is not well constrained, and the nonrelativistic matter $\Omega_{\rm m}$ has to be less than $0.45$ at about 3$\sigma$.  The contours in Fig.\ 1 are in reasonable accord with those shown in Fig.\ 2 of \citet{samushia08b} which were derived using the ``non-radial'' BAO peak scales measured by \citet{percivaletal07a} and \citet{eisensteinetal}. When the radial BAO data is combined with supernova data the resulting constraints are significantly stronger, see Fig.\ 5. The nonrelativistic matter density parameter is now constrained to be $0.15<\Omega_{\rm m}<0.35$ at about 3$\sigma$ confidence level while the cosmological constant density parameter lies in the $0.45<\Omega_\Lambda<0.9$ range. The best-fit values are close to the spatially-flat model. The constraints computed from the joint one-dimensional likelihoods shown in Figs.\ 8a and 8b are listed in Table\ 1. They are much more restrictive than those derived from the individual data sets, because those constrain combinations of dark energy parameters that are in some sense ``orthogonal'' in parameter space.

For the XCDM parametrization (for which we consider only spatially-flat models) the confidence level contours derived from radial BAO measurements are broad and a range of $\Omega_{\rm m}$ and $\omega_{\rm x}$ values are acceptable. The constraints are shown in Fig.\ 2. They are similar to the results shown in \citet[Fig.\ 1]{gaztanaga08b}. When we use the WMAP measured value of the ratio $l_s$ (thick lines) the nonrelativistic matter $\Omega_{\rm m}$ has to be less than $0.4$ at about 3$\sigma$. These confidence level contours are in reasonable accord with the ones shown in \citet[Fig.\ 3]{samushia08b} which were derived using the non-radial BAO data of \citet{percivaletal07a} and \citet{eisensteinetal}. However, compared to the non-radial BAO scale measurements the radial BAO scale measurements better constrain $\omega_{\rm x}$ from below and tend to favor higher values of it.  The joint constraints from radial BAO and supernova data are shown in Fig.\ 6. The joint likelihood constrains the equation of state parameter to be $-0.7<\omega_{\rm x}<-1.2$ at about 3$\sigma$ confidence, while $0.2<\Omega_{\rm m}<0.35$. For XCDM the joint constraints do not depend much on the method used for the analysis of the radial BAO data, unlike the $\Lambda$CDM and $\phi$CDM cases. One-dimensional joint likelihood functions for $\Omega_{\rm m}$ and $\omega_{\rm x}$ are shown in Figs.\ 8c and 8d and the corresponding constraints are given in Table\ 1.

The spatially-flat $\phi$CDM model confidence level contours are shown in Fig.~3. Here, the radial BAO measurements constrain $\Omega_{\rm m}$ to be between $0.15$ and $0.4$ at about 3$\sigma$, but the $\alpha$ parameter is not constrained well and large values of $\alpha$ (relatively rapidly evolving dark energy) are not ruled out, although the likelihood peaks at $\alpha=0$. These results are similar to the ones derived in  \citet[Fig.\ 4] {samushia08b} using the non-radial BAO peak scale measurements of \citet{percivaletal07a} and \citet{eisensteinetal}. Figure\ 6 shows type Ia supernova constraints on $\phi$CDM. From the Union data set alone, the nonrelativistic matter energy density parameter $\Omega_{\rm m}<0.4$, and if $\Omega_{\rm m}>0.1$, $\alpha<4.3$ at about 3$\sigma$ confidence. These contours are similar to and a little more constraining than the ones derived using the \citet{riess04} type Ia supernova Gold data \citep[Fig.\ 1]{wilson06}. When we combine the radial BAO measurements with the Union supernova data, Fig.\ 7, the $\alpha$ parameter is constrained to be less than 1.5 at 3$\sigma$ while $0.2<\Omega_{\rm m}<0.35$. One-dimensional joint likelihood functions for $\Omega_{\rm m}$ and $\alpha$ are shown in Figs.\ 8e and 8f and the corresponding constraints are listed in Table\ 1. For both radial BAO analysis methods $\Omega_{\rm m}$ is relatively well constrained and the likelihood for $\alpha$ peaks at low values of $\alpha$, close to the corresponding spatially-flat $\Lambda$CDM values.

From the analysis presented here and in \citet{samushia08b}, it is clear that neither the current radial or non-radial BAO peak scale measurements by themselves are able to tightly constrain cosmological parameters. However, given the shape of the resulting confidence contours, typically an elongated ellipse-like shape, current BAO data can provide very useful constraints when combined with other data like type Ia supernova measurements. As another example, the constraints derived from radial and non-radial BAO peak measurements complement the constraints derived from galaxy cluster gas mass fraction data \citep{allen08,samushia08a} and when used jointly they can tightly constrain parameters of dark energy models \citep{samushia08b}. It is interesting to note that for the $\Lambda$CDM and XCDM models the best fit values from the BAO and galaxy cluster gas mass fraction data sets are about 3$\sigma$ apart \citep[see][Figs.\ 2--3]{samushia08b}, but for the two data sets together the minimum $\chi^2$ is acceptable. This is, most probably, due to unknown systematic errors in one (or both) of the data sets and deserves further investigation.

It is also of interest to compare the constraints shown in Figs.\ 1--3 to those derived using other data. The radial BAO constraints shown in Figs.\ 1--3 are more restrictive than the ones derived from Hubble parameter versus redshift data \citep{samushia07} and from gravitational lensing data \citep{cha04}. However, they are comparable to the ones derived from earlier type Ia supernova data \citep{riess04} by \citet{wilson06}.

Current BAO peak scale data is sparse and can not tightly constrain dark energy parameters or differentiate between different dark energy models. In next few years a number of planned surveys like PAU \citep{benitez08}, BOSS, and WiggleZ \citep{glazebrook07} should measure the BAO peak scale with $1\%$ accuracy possibly up to a redshift of $z=1.2$. These future BAO measurements should prove to be of great significance in discriminating between different currently viable dark energy models and constraining cosmological parameters.

\acknowledgements
We acknowledge useful discussions with E.\ Gazta\~{n}aga and support from DOE grant DE-FG03-99EP41093 and INTAS grant 061000017-9258.

\begin{deluxetable}{ c c c c c }
\tablecaption{Best fit values and 1$\sigma$ range of $\Omega_{\rm m}$ and $p$.\tablenotemark{a}}
\tablehead{\colhead{model} & \colhead{$\Omega_{\rm m}$} & \colhead{$\Omega_{\rm m}\ 1\sigma \ {\rm range}$} & \colhead{$p$} & \colhead{$p \ 1\sigma \ {\rm range}$}}
\startdata
$\Lambda{\rm CDM}\quad ({\rm prior\ } l_s)$ & $0.27$          &$0.24 - 0.31$  &$\Omega_\Lambda=0.70$  &$0.64<\Omega_\Lambda<0.75$\\
$\Lambda{\rm CDM}\quad ({\rm prior\ } r_s)$&$0.25$          &$0.24 - 0.32$  &$\Omega_\Lambda=0.68$ &$0.62<\Omega_\Lambda<0.77$\\
${\rm XCDM}\quad ({\rm prior\ } l_s)$       &$0.24$          &$0.23 - 0.27$  &$\omega_{\rm x}=-0.91$&$-1.00<\omega_{\rm x}<-0.88$\\
${\rm XCDM}\quad ({\rm prior\ } r_s)$       &$0.25$          &$0.23 - 0.28$  &$\omega_{\rm x}=-0.93$&$-1.03<\omega_{\rm x}<-0.85$\\
$\phi{\rm CDM}\quad ({\rm prior\ } l_s)$       &$0.24$          &$0.22 - 0.26$  &$\alpha=0.27$&$0.02<\alpha<0.54$\\
$\phi{\rm CDM}\quad ({\rm prior\ } r_s)$       &$0.25$          &$0.24 - 0.27$  &$\alpha=0.20$&$\alpha<0.43$\\
\enddata
\tablenotetext{a}{Best-fit values of cosmological parameters $\Omega_{\rm m}$ and $p$, and 1$\sigma$ confidence level intervals, from one-dimensional likelihood functions derived from a joint analysis of radial BAO and type Ia supernova data. Entries labeled as ``prior $l_s$'' are computed using the WMAP measured ratio $l_s$ (and correspond to the thick lines in the figures). Entries labeled as ``prior $r_s$'' are computed using the WMAP measured value for the sound horizon at recombination $r_s$ (and correspond to the thin lines in the figures).}
\end{deluxetable}

\begin{figure}
\includegraphics[clip, trim=20mm 0 20mm 0, width=180mm, height=180mm]{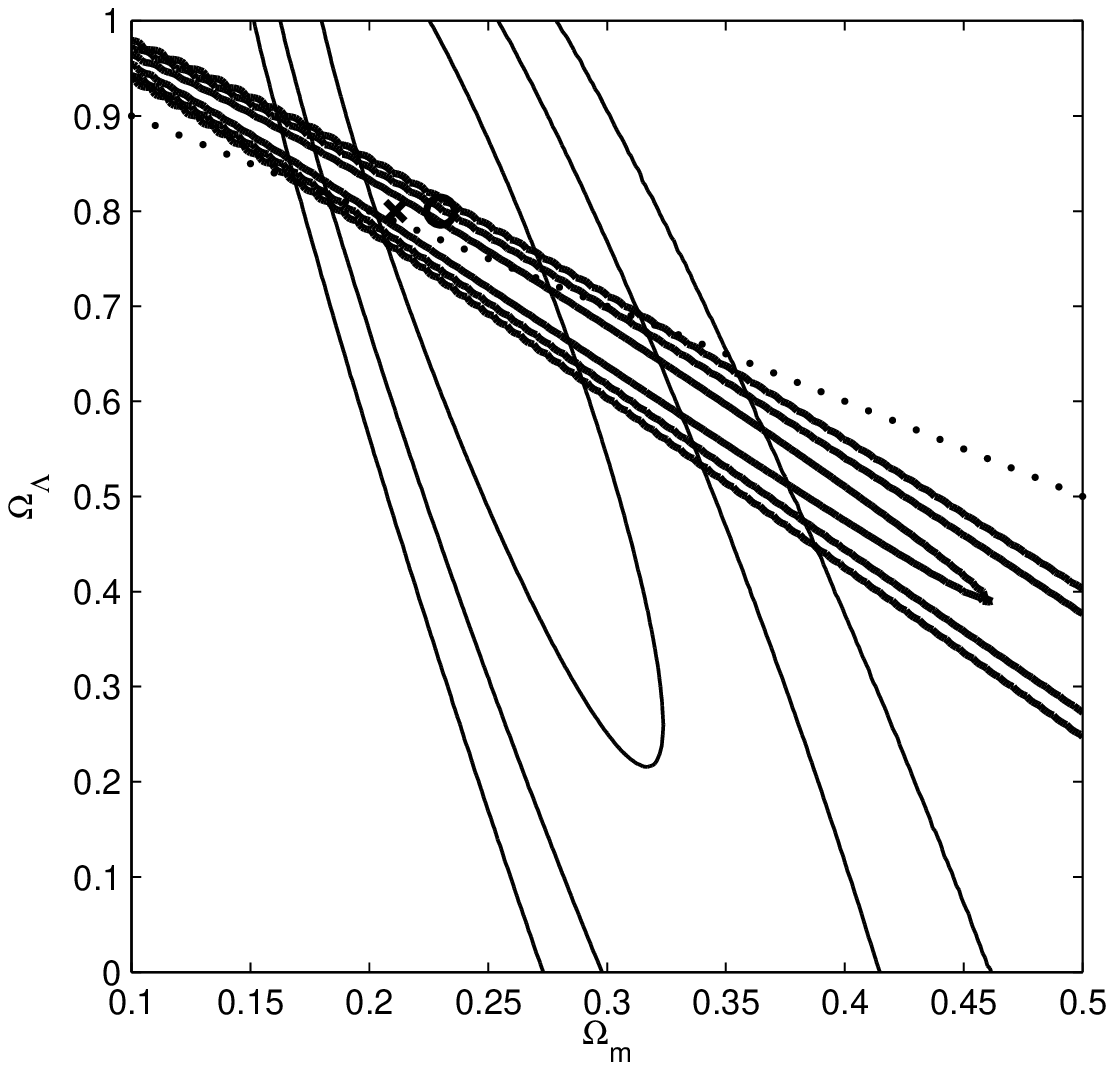}
\caption{1, 2, and 3$\sigma$ confidence level contours for the $\Lambda$CDM model from radial BAO measurements. Thick lines are derived using the WMAP measured ratio $l_s$ and the cross shows best-fit model parameters, $\Omega_{\rm m}=0.21$ and $\Omega_\Lambda=0.80$, derived from the two-dimensional likelihood function. The thin lines are derived using the WMAP value for the sound horizon at recombination $r_s$ and the circle shows best-fit model parameters, $\Omega_{\rm m}=0.23$ and $\Omega_\Lambda=0.80$, derived from the two-dimensional likelihood function. The dotted line demarcates spatially-flat $\Lambda$CDM models.}
\end{figure}

\begin{figure}
\includegraphics[clip, trim=20mm 0 20mm 0, width=180mm, height=180mm]{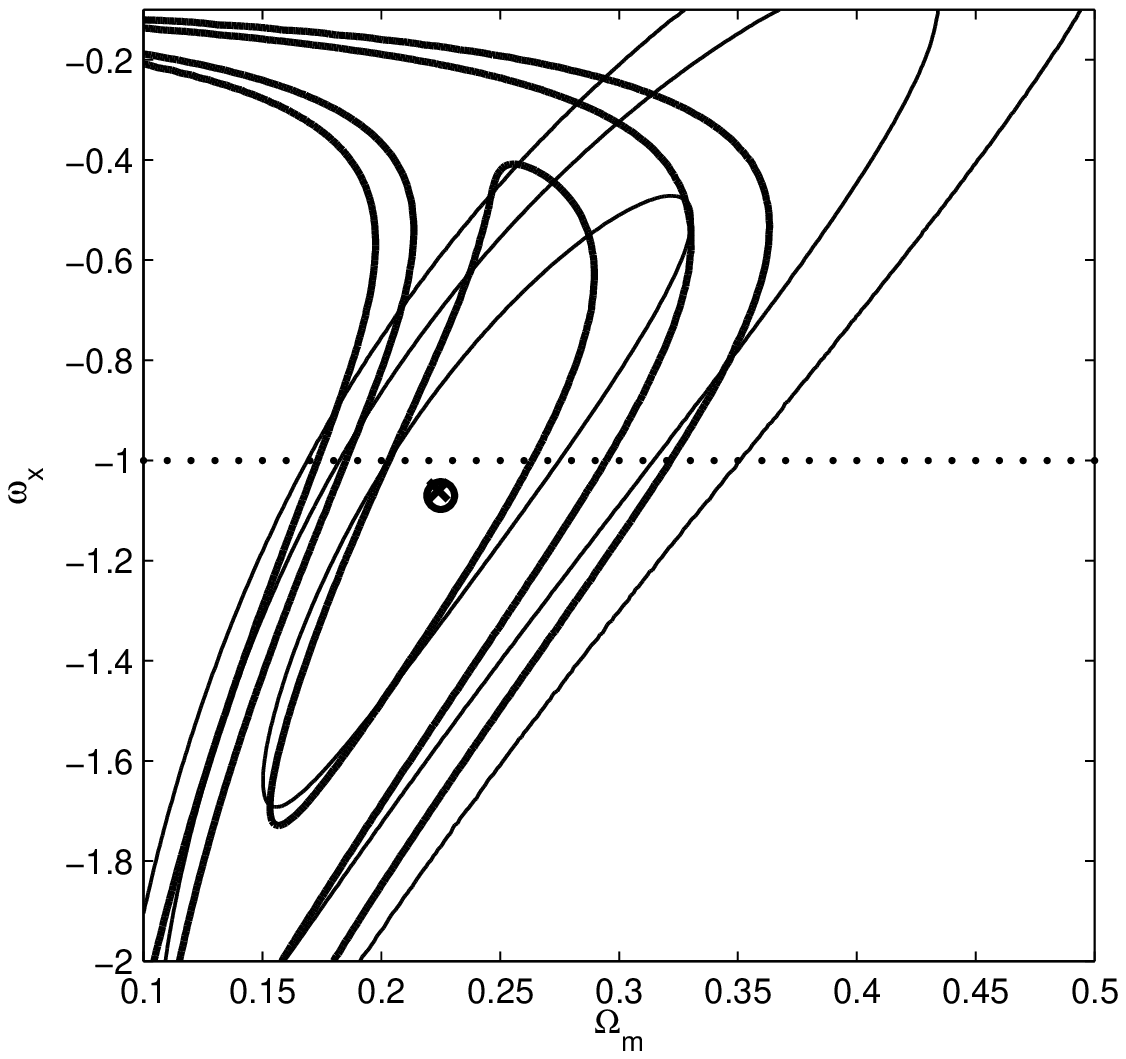}
\caption{1, 2, and 3$\sigma$ confidence level contours for the XCDM parametrization from radial BAO measurements. Thick lines are derived using the WMAP measured ratio $l_s$ and the cross shows best-fit model parameters, $\Omega_{\rm m}=0.22$ and $\omega_{\rm x}=-1.03$, derived from the two-dimensional likelihood function. The thin lines are derived using the WMAP value for the sound horizon at recombination $r_s$ and the circle shows best-fit model parameters, $\Omega_{\rm m}=0.22$ and $\Omega_{\rm x}=-1.03$, derived from the two-dimensional likelihood function. The dotted horizontal line demarcates spatially-flat $\Lambda$CDM models.}
\end{figure}

\begin{figure}
\includegraphics[clip, trim=20mm 0 20mm 0, width=180mm, height=180mm]{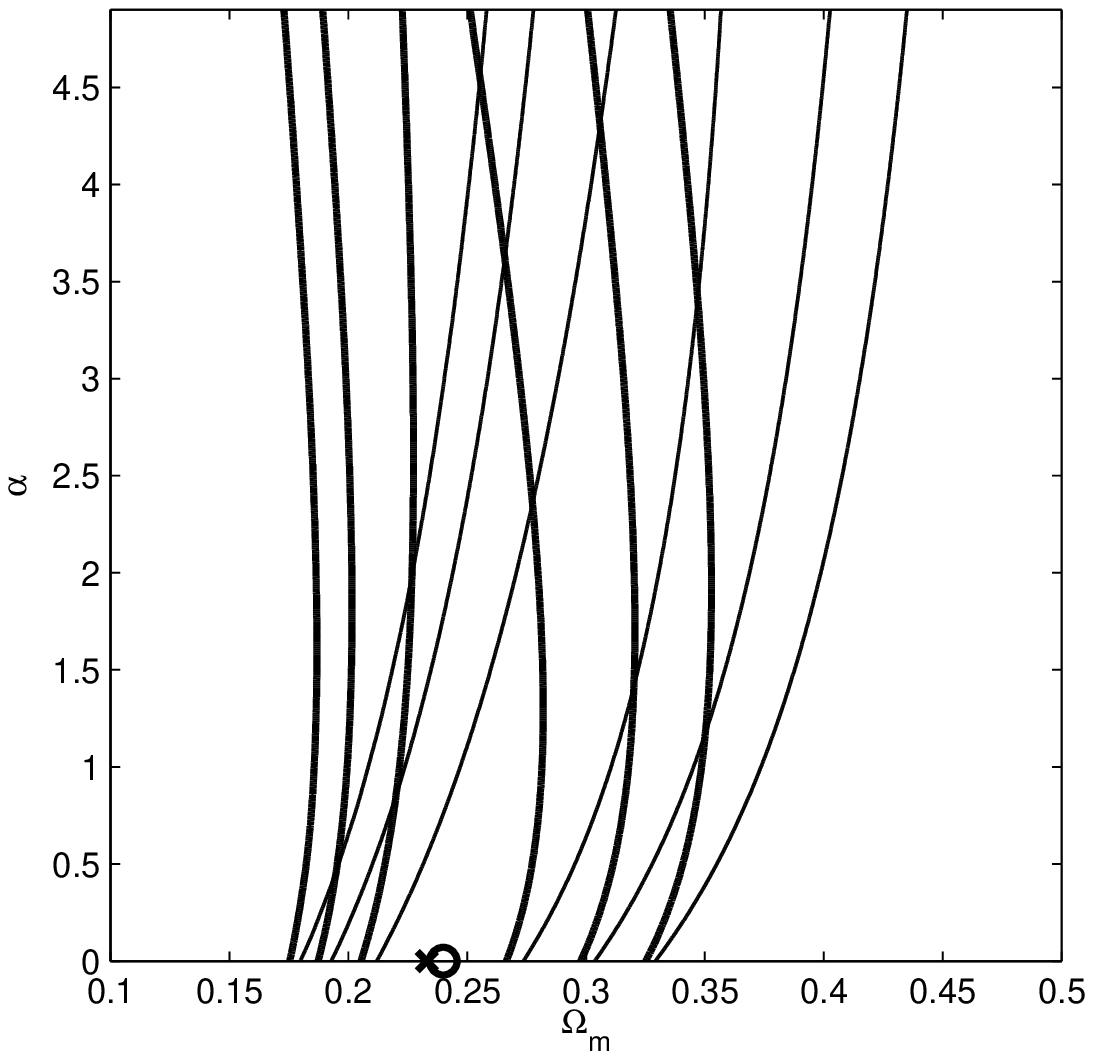}
\caption{1, 2, and 3$\sigma$ confidence level contours for the $\phi\rm{CDM}$ model from radial BAO measurements. Thick lines are derived using the WMAP measured ratio $l_s$ and the cross shows best-fit model parameters, $\Omega_{\rm m}=0.23$ and $\alpha=0.0$, derived from the two-dimensional likelihood function. The thin lines are derived using the WMAP value for the sound horizon at recombination $r_s$ and the circle shows best-fit model parameters, $\Omega_{\rm m}=0.24$ and $\alpha=0.0$, derived from the two-dimensional likelihood function. The horizontal axis with $\alpha=0$ corresponds to spatially-flat $\Lambda$CDM models.}
\end{figure}

\begin{figure}
\includegraphics[clip, trim=20mm 0 20mm 0, width=180mm, height=180mm]{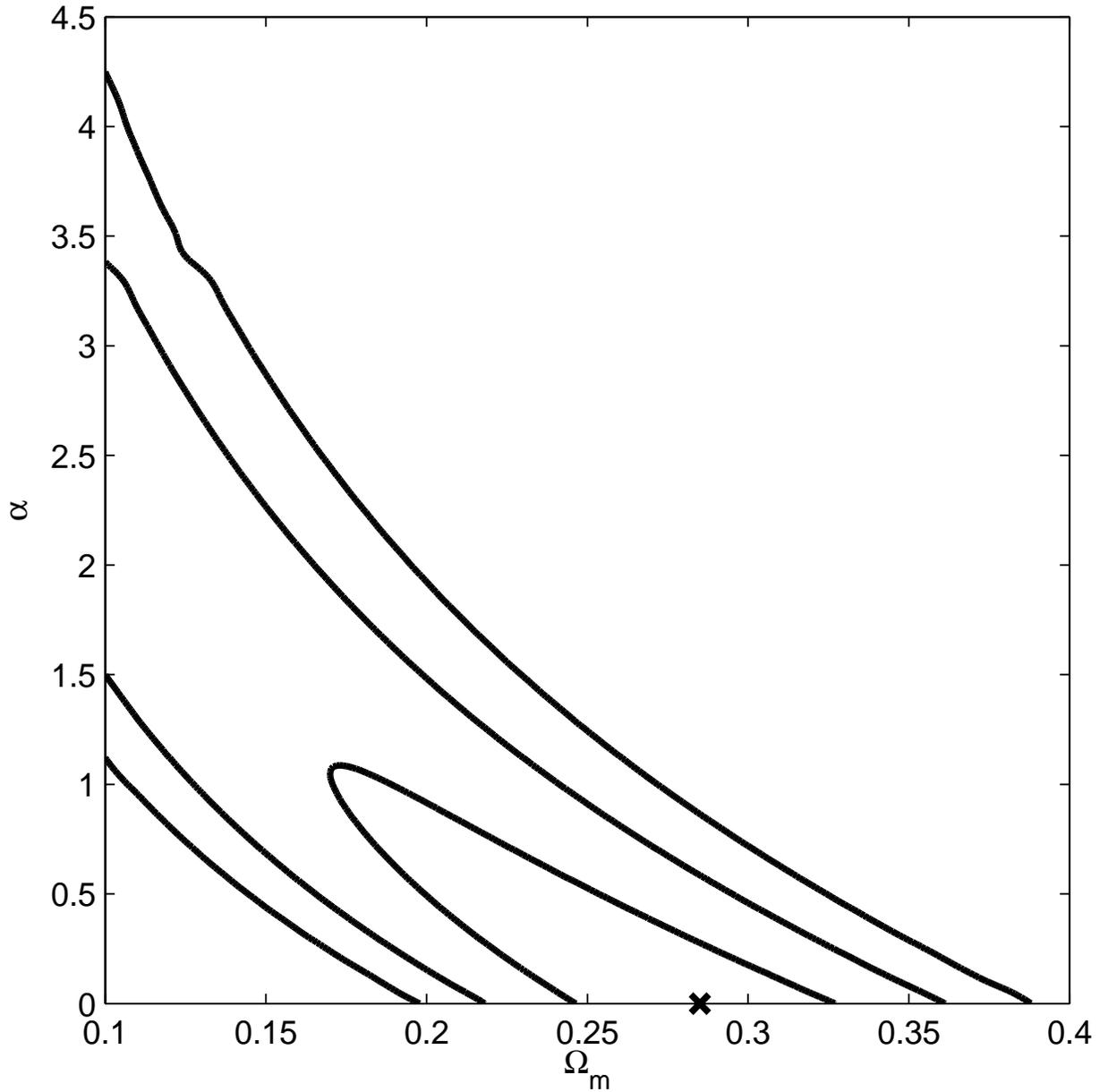}
\caption{1, 2, and 3$\sigma$ confidence level contours for the $\phi$CDM model from the supernova data. The cross shows best-fit model parameters, $\Omega_{\rm m}=0.29$ and $\alpha=0.0$, derived from the two-dimensional likelihood function. The horizontal axis with $\alpha=0$ corresponds to spatially-flat $\Lambda$CDM models.}
\end{figure}

\begin{figure}
\includegraphics[clip, trim=20mm 0 20mm 0, width=180mm, height=180mm]{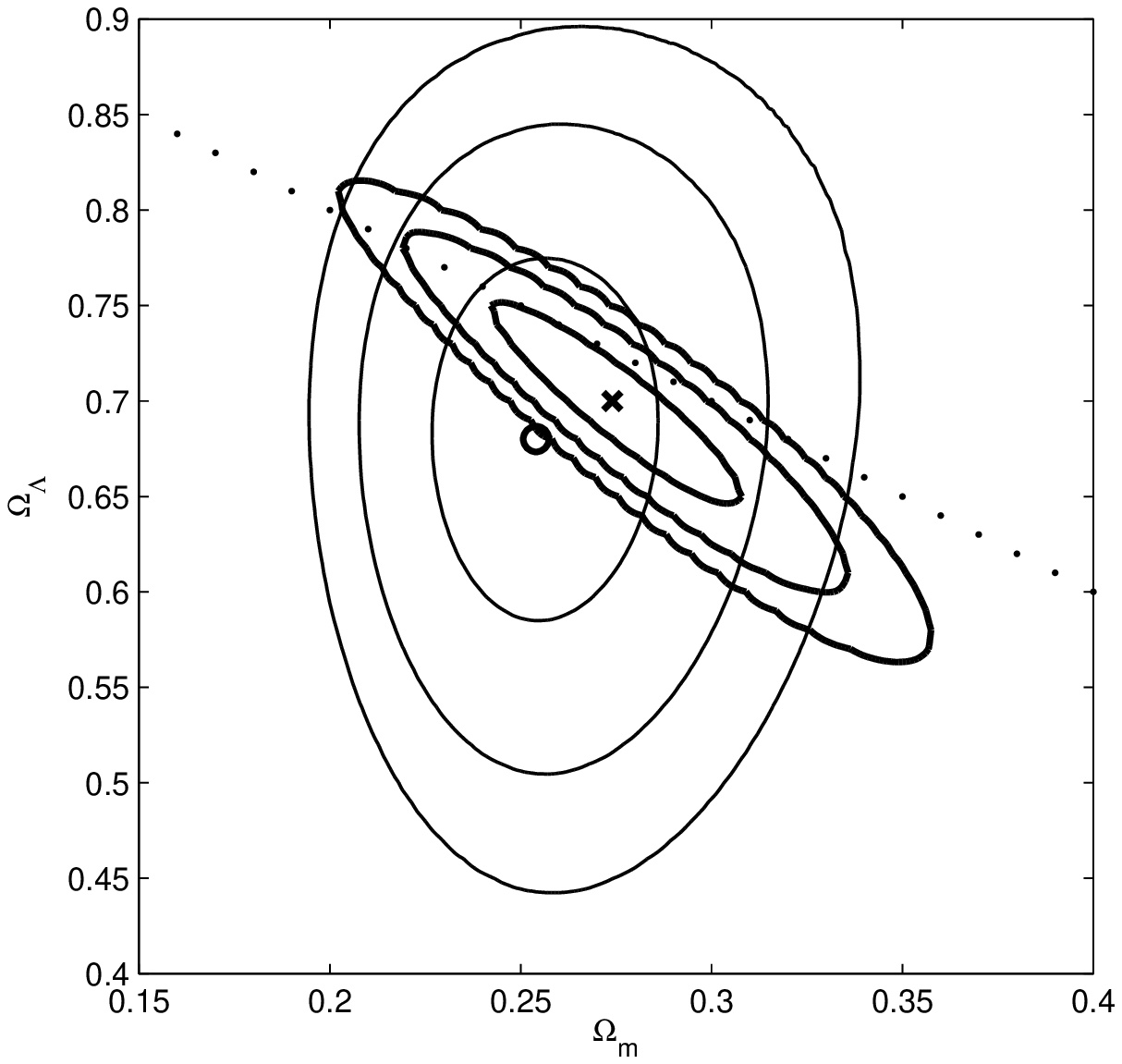}
\caption{1, 2, and 3$\sigma$ confidence level contours for the $\Lambda$CDM model derived using the joint likelihood from radial BAO and type Ia supernova measurements. Thick lines are derived using the WMAP measured ratio $l_s$ and the cross shows best-fit model parameters, $\Omega_{\rm m}=0.27$ and $\Omega_\Lambda=0.70$, derived from the two-dimensional likelihood function. The thin lines are derived using the WMAP value for the sound horizon at recombination $r_s$ and the circle shows best-fit model parameters, $\Omega_{\rm m}=0.25$ and $\Omega_\Lambda=0.68$, derived from the two-dimensional likelihood function. The dotted line demarcates spatially-flat $\Lambda$CDM models. The $\Omega_{\rm m}$ and $\Omega_\Lambda$ ranges shown here are smaller than those shown in Fig.\ 1.}
\end{figure}

\begin{figure}
\includegraphics[clip, trim=20mm 0 20mm 0, width=180mm, height=180mm]{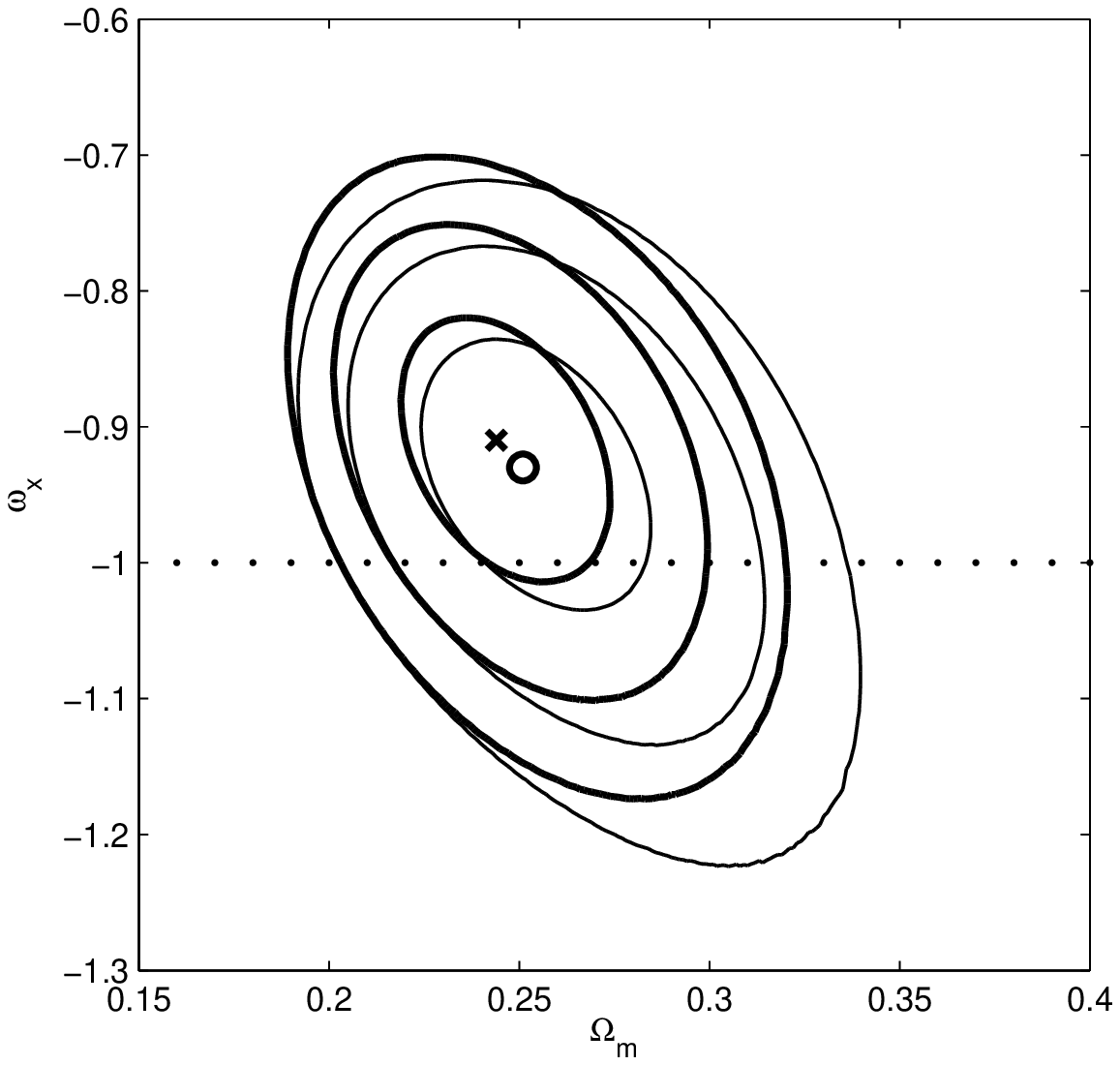}
\caption{1, 2, and 3$\sigma$ confidence level contours for the XCDM parametrization derived using the joint likelihood from radial BAO and type Ia supernova measurements. Thick lines are derived using the WMAP measured ratio $l_s$ and the cross shows best-fit model parameters, $\Omega_{\rm m}=0.24$ and $\omega_{\rm x}=-0.91$, derived from the two-dimensional likelihood function. The thin lines are derived using the WMAP value for the sound horizon at recombination $r_s$ and the circle shows best-fit model parameters, $\Omega_{\rm m}=0.25$ and $\Omega_{\rm x}=-0.93$, derived from the two-dimensional likelihood function. The dotted horizontal line demarcates spatially-flat $\Lambda$CDM models. The $\Omega_{\rm m}$ and $\omega_{\rm x}$ ranges shown here are smaller than those shown in Fig.\ 2.}
\end{figure}

\begin{figure}
\includegraphics[clip, trim=20mm 0 20mm 0, width=180mm, height=180mm]{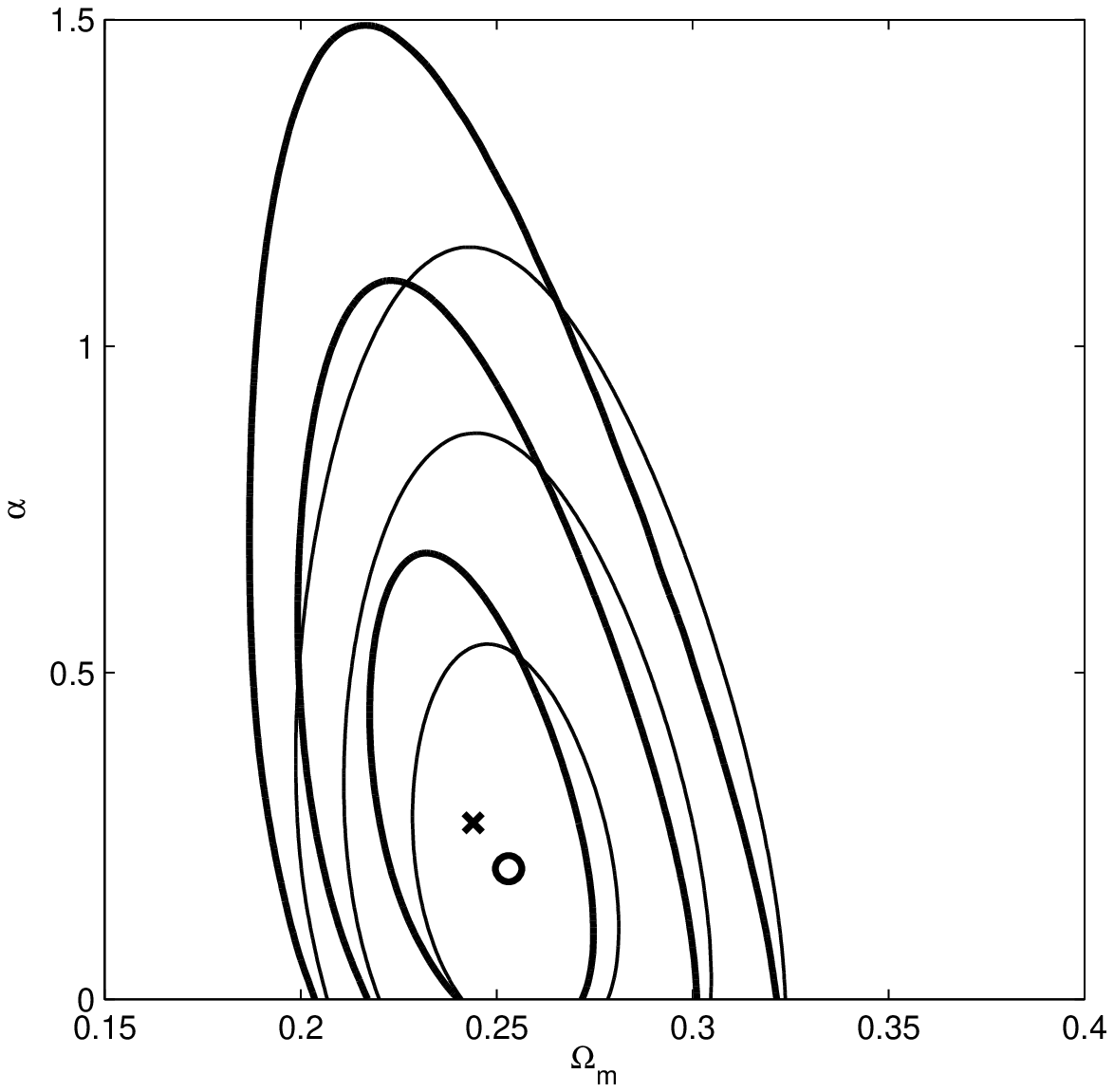}
\caption{1, 2, and 3$\sigma$ confidence level contours for the $\phi\rm{CDM}$ model derived using the joint likelihood from radial BAO and type Ia supernova measurements. Thick lines are derived using the WMAP measured ratio $l_s$ and the cross shows best-fit model parameters, $\Omega_{\rm m}=0.24$ and $\alpha=0.27$, derived from the two-dimensional likelihood function. The thin lines are derived using the WMAP value for the sound horizon at recombination $r_s$ and the circle shows best-fit model parameters, $\Omega_{\rm m}=0.25$ and $\alpha=0.20$, derived from the two-dimensional likelihood function. The horizontal axis with $\alpha=0$ corresponds to spatially-flat $\Lambda$CDM models. The $\Omega_{\rm m}$ and $\alpha$ ranges shown here are smaller than those shown in Figs.\ 3 and 4.}
\end{figure}

\begin{figure}
\includegraphics[clip, trim=10mm 0 10mm 0, width=180mm, height=180mm]{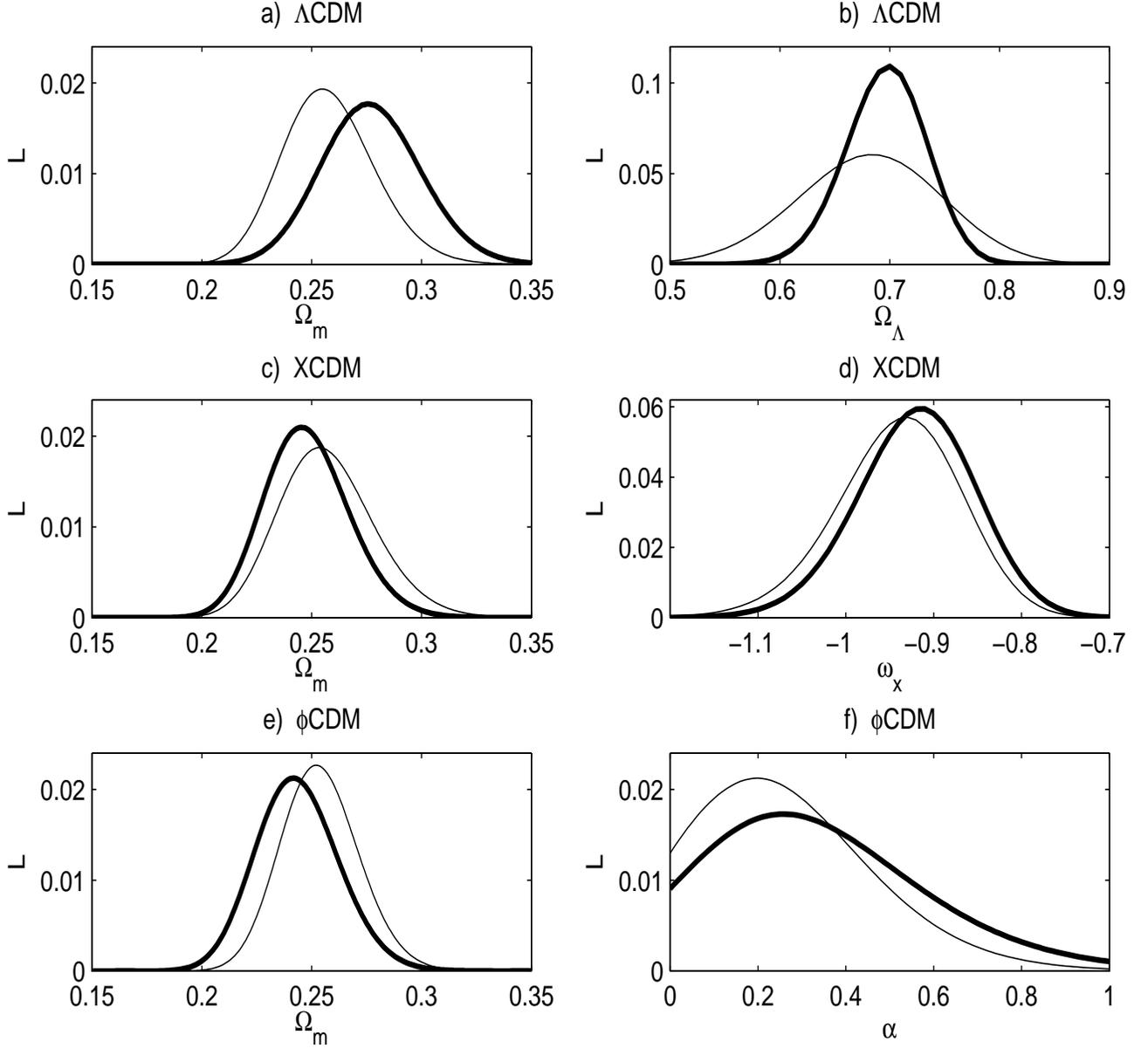}
\caption{One-dimensional likelihood functions for cosmological parameters for the three models derived using the joint likelihood from radial BAO and type Ia supernova data. Thick lines show results derived using the WMAP measured ratio $l_s$ while thin lines are derived using the WMAP value for the sound horizon at recombination $r_s$. In each of the six subplots total likelihoods are normalized to one.}

\end{figure}
\end{document}